# Forecasting per-patient dosimetric benefit from daily online adaptive radiotherapy for cervical cancer

**Rupesh Ghimire[1], Kevin L. Moore[1], Daniela Branco[1], Dominique L. Rash[1], Jyoti Mayadev[1], and Xenia Ray[1]**

## Abstract

Objective: Adaptive Radiotherapy (ART) is an emerging technique for treating cancer patients which facilitates higher delivery accuracy and has the potential to reduce toxicity. However, ART is also resource-intensive, requiring extra human and machine time compared to standard treatment methods. In this analysis, we sought to predict the subset of node-negative cervical cancer patients with the greatest benefit from ART, so resources might be properly allocated to the highest-yield patients.

Approach: CT images, initial plan data, and on-treatment Cone-Beam CT (CBCT) images for 20 retrospective cervical cancer patients were used to simulate doses from daily non-adaptive and adaptive techniques. We evaluated the correlation ($R^2$) between dose and volume metrics from initial treatment plans and the dosimetric benefits to the $Bowel\ V_{40Gy}$, $Bowel\ V_{45Gy}$, $Bladder\ D_{mean}$, and $Rectum\ D_{mean}$ from adaptive radiotherapy using reduced 3mm or 5mm CTV-to-PTV margins. The LASSO technique was used to identify the most predictive metrics for $Bowel\ V_{40Gy}$. The three highest performing metrics were used to build multivariate models with leave-one-out validation for $Bowel\ V_{40Gy}$.

Main Results: Patients with higher initial bowel doses were correlated with the largest decreases in Bowel $V_{40Gy}$ from daily adaptation (linear best fit $R^2$=0.77 for a 3mm PTV margin and $R^2$=0.8 for a 5mm PTV margin). Other metrics had intermediate or no correlation. Selected covariates for the multivariate model were differences in the initial $Bowel\ V_{40Gy}$ and $Bladder\ D_{mean}$ using standard versus reduced margins and the initial bladder volume. Leave-one-out validation had an $R^2$ of 0.66 between predicted and true adaptive $Bowel\ V_{40Gy}$ benefits for both margins.

Significance: The resulting models could be used to prospectively triage cervical cancer patients on or off daily adaptation to optimally manage clinical resources. Additionally, this work presents a critical foundation for predicting benefits from daily adaptation that can be extended to other patient cohorts.

## 1. Introduction

External beam radiotherapy delivered daily over 5-6 weeks is a standard treatment technique for cervical cancers which affect 604,000 women a year and results in over 342,000 deaths worldwide[1]. The goal of radiotherapy is to deliver a therapeutic dose to the tumour while minimizing the exposure to the surrounding tissues. To achieve this trade-off, targets and organs-at-risk (OARs) are delineated on a simulation computed tomography (CT) and highly optimized plans are created using Intensity Modulated Radiotherapy (IMRT) or Volumetric Modulated Arc Therapy (VMAT) techniques. Targets include the upper third or half of the vagina, cervix, uterus, parametria, and pelvic nodes. However, when the patient arrives for treatment each day, these targets may have changed shape or position, meaning the plan can be suboptimal for that day's internal anatomy. In particular, the uterus has been measured to move substantially between treatments by as much as 4-6 cm[2], [3]. To account for these changes, and ensure the targets always receive the prescribed daily dose, large margins of 1-2cm are added to the targets during initial planning [4], [5]. This approach comes at the expense of delivering higher doses to the nearby organs that overlap with the margin area, which in turn can result in toxicities. These toxicities range from acute to chronic, sometimes lasting years from the date of treatment. Irradiation of the bowel and rectum volume may cause gastrointestinal (GI) toxicities like diarrhoea, cramping and abdominal discomfort[6]–[8]. In RTOG 1203, greater than 1/3 of patients reported GI toxicity of acute frequent diarrhoea with IMRT for cervical or endometrial cancer. Other common symptoms include intolerance to certain food types, malabsorption, intermittent diarrhoea, and bladder problems. In addition to the risks from excess normal tissue dose, the anatomical changes[9] of the target structures can also be larger than these margins, resulting in delivering lower than prescribed doses to portions of the target and thus potentially decreasing local control.

Adaptive radiotherapy (ART), initially proposed by Yan et.al[10], is a unique approach for modifying the treatment plan by systematically monitoring treatment variations and incorporating the feedback to re-optimize the original plan. The application of adaptive radiotherapy for cervical cancer has long been of interest as a strategy for minimizing margins and decreasing dose to normal tissues while still ensuring excellent target dosing. Multiple studies have indicated dosimetric benefits to patients from various adaptive strategies including a library-of-plans technique [11], [12], where multiple plans are created ahead of treatment and the one most representative of each day's bladder filling is selected at treatment. Newer commercial solutions, such as the Varian Ethos[13], Elekta Unity MR-Linac[14], and ViewRay MRIdian[15] have allowed clinics to implement daily online ART[16] where a new plan is fully re-optimized at every treatment based on images acquired that day with the patient in the treatment position. Recent studies by Branco *et al* and Yock *et al* demonstrated that cone-beam CT (CBCT) based ART, facilitated by the Varian Ethos could be implemented for cervical cancer within a total average treatment time slot of 25 minutes[17], [18]. Their studies found it would increase the volume of the target receiving 99% and 95% of the prescription dose respectively and decrease dose to the bowel, rectum, and bladder.

Despite its advantages, online ART is resource-intensive as it requires the presence of a trained expert to contour the daily anatomy at each treatment procedure, extra online or offline supervision by the physician to review these contours, and lengthens the overall appointment times. These longer treatment times can also be challenging for patients who must remain still for the duration and may be in an uncomfortable position. Strategies to mitigate these issues have included training therapists[19] or physicists[17] to adjust targets at each fraction in a timely manner and evaluating whether auto-segmentation results are robust enough to minimize manual edits and decrease the total on-treatment time[20]. An alternative strategy is to focus efforts on identifying patients most likely to benefit from ART and target them with this treatment method. This would be valuable for cervical cancer patients as existing data indicates certain patients have more anatomical motion or unfavourable anatomy which may result in greater benefit from ART [20]. Additionally, the benefit from ART for all cervical cancer patients can be enhanced by using reduced target margins, though the magnitude of this additional benefit per patient is likely to differ based on the location and daily motion of their OARs.

The purpose of this study was to evaluate the net dosimetric impact of ART with different reduced margins on cervical cancer patients and then determine if the patient-specific benefit can be predicted from metrics available on the initial simulation CT.

## 2. Methods

All data in this study was performed under the approval of our institutional review board, XXXXXXXX. For this study we collected retrospective images (initial planning CTs and weekly CBCTs) and planning data from 20 cervical cancer patients previously treated at our institution with external beam radiation (45 Gy in 25 fractions). To delineate the targets for their initial clinical plans, patients were imaged with both a full and empty bladder using our GE Discovery CT590 RT Simulator. All targets were drawn by an attending physician. Clinical target volumes follow consensus guidelines[21] and consisted of CTV1 (gross tumour, cervix, and uterus), CTV2 (parametria and upper third of the vagina or upper half to two-thirds of the vagina if it was clinically involved), and CTV3 (common, external, and internal iliac and presacral lymph nodes). All CTVs were drawn on the CT image used for planning (either the full or empty scan at physician discretion). CTV1 was drawn on both the full and empty scans and the union of these two structures was used to create an ITV. Further margins were added to this ITV, CTV1,

CTV2, and CTV3 to create the non-adaptive "standard-of-care" (SOC) $PTV_{SOC}$. Specifically, 7mm symmetric margins are added to the ITV, 15mm to CTV1, 10mm to CTV2, and 5mm to CTV3. Clinically treated plans were created in Eclipse version 16.1.0 and were optimized using volumetric modulated arc therapy (VMAT) and normalized so that $PTV_{SOC}$ $D_{95\%}$ was equal to 100% of the 45 Gy prescription dose. These plans were used to obtain the patient's initial dose metrics and to calculate the daily delivered dose without adaptation. This process is illustrated in Figure 1 and described in detail below.

To create the adapted dose data, we used a research version of the Ethos clinical system. This includes a non-clinical Ethos Treatment Planning System (TPS) for creating the plans that will be adapted and an emulator of the Ethos treatment machine, so that the entire end-to-end adaptive process can be performed on retrospective CBCT images. The emulator uses the same optimization engine, dose calculation algorithms, and auto-segmentation algorithms as a clinical Ethos v1.1 thus allowing us to mock-treat online adaptive sessions using retrospective CBCTs and obtain the same adapted plans we would calculate on a clinical system. For the emulator to optimize an adapted plan, it needs an initial Ethos-generated plan with a prioritized list of dose goals. The CT images used for the initial clinical plan in Eclipse along with the clinically used structure set were imported into the Ethos TPS to create these new plans that could be adapted at each fraction. Because adaptation improves setup accuracy, a new PTV with smaller 3mm margins on CTV1, CTV2, and CTV3 was created for each patient using 3mm margins. Note, no ITV was created for CTV1 as the adapted plans would inherently account for daily variations in bladder filling. A new plan was optimized using this $PTV_{3mm}$ and the Ethos TPS for 45 Gy in 25 fractions. Plans were normalized so $PTV_{3mm}$ $D_{95\%}$ were equal to 100% of the prescription dose. The Ethos generated plans used either 9- or 12-field Intensity Modulated Radiation Therapy (IMRT) beam arrangements. This was done because (1) studies have shown that IMRT plans optimized by the Ethos Emulator out-perform VMAT plans[22], (2) IMRT plans[23] optimize faster than VMAT plans (2 vs 7 minutes) and this extra speed can be helpful for adaptive radiotherapy where the patient must lie on the couch the entire time a new plan is created, and (3) in our clinical practice IMRT plans are used for daily adaptation of conventionally fractionated patients for the prior two reasons. All Ethos-generated plans were evaluated for acceptability against our standard clinical metrics prior to simulating the adaptive treatments. If found unacceptable, dose goals were adjusted until a clinically acceptable plan was produced.

The next step was generating simulated daily adapted plans in the Ethos treatment machine emulator. For each of the 20 patients, we simulated 5 adapted treatments using kV-CBCT images evenly spaced throughout their treatment (i.e. fractions 5, 10, 15, 20, and 25). The emulator process has been described in previous work by multiple groups[18], [20] and is summarized here in brief. When imported into the Ethos emulator, the CBCT images are auto segmented to delineate the influencer structures. For cervical cancer targets, these influencer structures are the bladder, bowel, rectum, and uterus. Any necessary edits to this auto-segmentation were made to ensure accurate structures for optimization. Once the influencers are approved, the target structures are auto-segmented and then correspondingly hand-edited and approved to create the adaptive plan (Daily$_{ADP}$). The adaptive plan is optimized using the same planning goals as the initial Ethos plan, and the adapted dose is calculated onto a synthetic CT[24], [25] (a deformed version of the simulation CT based on that fraction's CBCT) which is also auto-generated by the Ethos adaptive software.

All structure editing was performed by a trained group of physicists which mimics our clinical adaptive workflow. All edited targets were reviewed, edited (if necessary), and approved by an experienced radiation oncologist (MD) specializing in gynaecologic cancers.

# Standard of Care and Adapted Plan Generation in this Study

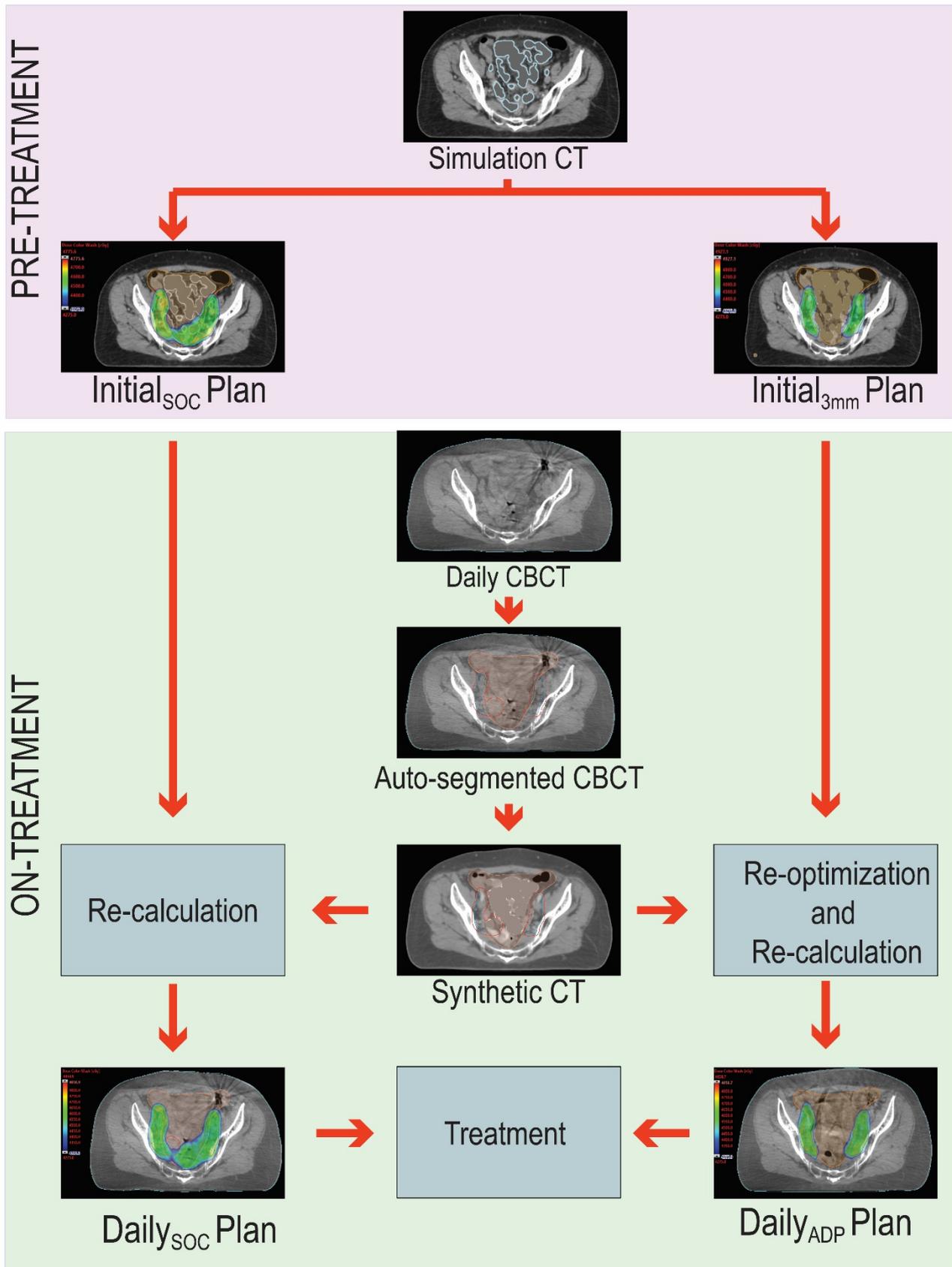

The initial clinical plan that had been treated for these patients was also recalculated on the synthetic CTs to determine the non-adaptive delivered doses. This was done in Eclipse v.16.1.0 by auto-registering the synthetic CTs to the initial planning CT based on bony anatomy and then adjusting as necessary to maximize target alignment as would be performed by radiation therapy technicians during a standard image-guided radiotherapy (IGRT) workflow. Because the Ethos machine has only 3 translational degrees of freedom for the couch, these IGRT registrations used only longitudinal, lateral, and vertical shifts.

Based on the process depicted in Figure 1, each patient had 12 calculated dose distributions: their initial clinically treated plan that used standard margins on the planning CT ($Initial_{SOC}$), their initial Ethos-generated plan that used 3mm margins on the planning CT ($Initial_{3mm}$), their $Initial_{SOC}$ plan recalculated on the 5 synthetic CTs using the anatomy-of-the-day from the 5 weekly CBCTs ($Daily_{SOC}$), and 5 fully reoptimized plans for the anatomy-of-the-day from the 5 weekly CBCTS ($Daily_{ADP,3mm}$).

Additionally, to evaluate the effect of using an intermediate margin decrease, we used a 5mm margin on CTV1-3 to create $PTV_{5mm}$, and then we reoptimized to obtain $Initial_{5mm}$ and $Daily_{ADP,5mm}$ plans. To ensure the exact same daily CTVs and OARs were used, these plans were reoptimized using a clinically validated RapidPlan™ model for cervical cancer[26]–[29] in the Eclipse TPS. This step was performed in Eclipse rather than repeating the adaptive process using the Ethos emulator because the emulator would have required re-contouring the targets and OARs at each fraction and thus led to small variations in volumes. This process created six new plans: an $Initial_{5mm}$ and five $Daily_{ADP,5mm}$ resulting in a total of 18 calculated dose distributions per patient.

Dose metrics were extracted from these plans and analysed as described below:

## 2.1 Dose Metrics

Dose volume histogram (DVH) metrics were obtained for the bowel, bladder, and rectum structures. For all plans, the bowel structure consisted of segmented bowel loops instead of a generic peritoneal space (bowel space) structure.

We measured the volume for each structure, the overlap volume of each structure with each of the three PTVs ($PTV_{SOC}$, $PTV_{5mm}$, and $PTV_{3mm}$), and the following dose metrics from each plan: $Bowel\ V_{40Gy}$ (cc), $Bowel\ V_{45Gy}$(cc), $Bladder\ D_{mean}$(Gy), $Bladder\ V_{45Gy}$(cc), $Rectum\ D_{mean}$(Gy), $Rectum\ V_{35Gy}$(cc) and $Rectum\ V_{40Gy}$(cc). The improvement from margin reduction on the initial plans was calculated by subtracting $Initial_{3mm}$ from $Initial_{SOC}$ plan and is represented by $Initial_{SOC-3mm}$. We measured the improvement from both margin reduction and daily adaptation, $\Delta_{SOC-ADP,3mm}$ by calculating the difference in $Daily_{SOC}$ and $Daily_{ADP,3mm}$ for each dose metric and averaging this difference over the five fractions for each patient. The mean of the daily plans across the five fractions are denoted by $\overline{Daily_{SOC}}$ and $\overline{Daily_{ADP,3mm}}$ respectively. These analyses were fully repeated for the 5mm reduced margin.

## 2.2 Effect of Margin Reduction and Adaptation

The impact of not adapting was tested using Wilcox signed-rank test for the $Initial_{SOC}$ versus $Daily_{SOC}$ metrics. The impact of the two margin reduction schemas were evaluated by using one-sided Wilcox ranked-sign test to compare dose metrics from $Initial_{SOC}$ to $Initial_{3mm}$ and $Initial_{5mm}$ metrics respectively. The impact of margin reduction combined with daily adaptation was assessed by using one-sided Wilcox ranked-sign tests[30] to compare the dose metrics from $\overline{Daily_{SOC}}$ to $\overline{Daily_{ADP,3mm}}$ and $\overline{Daily_{ADP,5mm}}$. P-values for these tests were corrected for multiplicity testing using the Bonferroni method[31].

## 2.3 Univariate relationships

To predict a patient's benefit from adaptation, we evaluated the linear relationship between changes in dose metrics measured from $Initial_{SOC}$ and $Initial_{3mm}$ plans and the observed improvement with adaptation $\Delta_{SOC-ADP,3mm}$ for four metrics: $Bowel\ V_{40Gy}$ (cc), $Bowel\ V_{45Gy}$(cc), $Bladder\ D_{mean}$(Gy), and $Rectum\ D_{mean}$(Gy). The $Initial_{SOC-3mm}$ difference is an estimate of the benefit from just the margin reduction facilitated by adaptation. We hypothesized this effect would dominate the overall benefit from adaptation. This analysis was repeated for the 5mm PTV margin. We also evaluated if the initial

Bladder, Rectum, and Bowel volumes were related to $\Delta_{SOC-ADP,3mm}$ and if the overlap of these structures with the various PTVs (PTV$_{SOC}$, PTV$_{5mm}$, and PTV$_{3mm}$) on the initial CT were related to $\Delta_{SOC-ADP,3mm}$. For each combination the R-squared ($R^2$) statistical value[32] was evaluated where >0.7 was considered strong correlation, 0.3-0.6 was considered moderate correlation, and <0.3 was considered a weak correlation[33].

### 2.4 Multivariate Modelling

Higher values of $Bowel\ V_{40Gy}$ (cc) have been associated with increased gastrointestinal toxicity [34] for cervical cancer patients. Thus, a patient likely to have a large decrease in this metric from adaptation may have a substantial improvement in outcomes and could be a preferred candidate for daily adaptation. To best identify these patients, we built a multivariate model to predict $Bowel\ V_{40Gy}\ \Delta_{SOC-ADP,3mm}$ from the initial planning data. To determine the most significant predictors for the multivariate model, we implemented multivariate linear regression using the least absolute shrinkage and selection operator (LASSO)[35] technique. Possible covariates included in the LASSO were all dose differences and volume metrics that were analysed in the univariate analysis that had an $R^2$>0.2. The three most predictive covariates obtained from the shrinkage in the LASSO were selected for the multivariate model. We then used leave-one-out validation to evaluate the model performance. In this technique the model parameters are fit using all but one patient and then a prediction is created for the left-out patient. The process is repeated, re-fitting the model on each iteration until a prediction has been made for each patient. We then evaluated the correlation between the predicted and true $Bowel\ V_{40Gy}\ \Delta_{SOC-ADP,3mm}$ by calculating the $R^2$. This analysis was repeated for the 5mm PTV margin.

## 3. Results

### 3.1 Effect of Margin Reduction and Adaptation

In this section, we report the population-wide effect on the dose metrics with and without adaptation. In the supplementary data, S. Table 1, we list the average and standard deviation of the OAR metrics across the 20 patients calculated from the initial and daily plans for each margin (SOC, 5mm, and 3mm). Table 1 below shows the mean value of the differences and the results of the Wilcoxon test for 8 metrics corresponding to the paired plans with and without adaptation using either of the two different margins: $Initial_{SOC}$ vs $Initial_{5mm}$, $Initial_{SOC}$ Vs $Initial_{3mm}$, $Initial_{SOC}$ vs $\overline{Daily}_{SOC}$, $\overline{Daily}_{SOC}$ vs $\overline{Daily}_{ADP,5mm}$, and $\overline{Daily}_{SOC}$ vs $\overline{Daily}_{ADP,3mm}$. As expected, there was no significant difference for the $Initial_{SOC}$ versus $\overline{Daily}_{SOC}$. The comparisons between $Initial_{SOC}$ and $Initial_{margin}$ as well as $Daily_{SOC}$ versus $Daily_{ADP}$ using either 3mm or 5mm margins all resulted in corrected p-values well below 0.05, suggesting a significant difference from margin reduction alone and from adaptation with margin reduction. Calculated p-values are available in Supplementary S. Table 2.

Figure 2 shows the OAR metric values from the $Initial_{SOC}$, $\overline{Daily}_{SOC}$, $\overline{Daily}_{ADP,5mm}$, and $\overline{Daily}_{ADP,3mm}$ plans for the four metrics we sought to predict. Plotted values for the Initial$_{5mm}$ and Initial$_{3mm}$ compared to Initial$_{SOC}$ values are available in the Supplementary Data (S. Figure 1). Each patient's values are represented by a different colour line. Changes from $Initial_{SOC}$ to $\overline{Daily}_{SOC}$ represent variation in planned versus delivered metrics due to anatomical changes. Systematic decreases were expected for $\overline{Daily}_{ADP,5mm}$ and $\overline{Daily}_{ADP,3mm}$ compared to $\overline{Daily}_{ADP}$ and were observed for most patients for all metrics. These represent the benefit in adaptive radiotherapy using reduced margins. In a few instances the $\overline{Daily}_{SOC}$ plan had lower values than the $Daily_{ADP,5mm}$ plan (e.g. yellow-green patient values for $Bladder\ D_{mean}$). This was observed for low $Initial_{SOC}$ dose values and may result from differences in planning techniques (VMAT versus IMRT) or unfavourable daily changes in anatomy (e.g. smaller bladder at treatment than planning). Additionally increases in doses to OARs with adaptive plans may be balanced by increased doses to the target volumes.

| Metrics | $Initial_{SOC}$ - $Initial_{5mm}$ ($\mu \pm \sigma$) | $Initial_{SOC}$ - $Initial_{3mm}$ ($\mu \pm \sigma$) | $Initial_{SOC}$ - $\overline{Daily_{SOC}}$ ($\mu \pm \sigma$) | $\overline{Daily_{SOC}}$ - $\overline{Daily_{ADP,5mm}}$ ($\mu \pm \sigma$) | $\overline{Daily_{SOC}}$ - $\overline{Daily_{ADP,3mm}}$ ($\mu \pm \sigma$) |
|---|---|---|---|---|---|
| $Bladder\ Dmean$(Gy) | $6.94 \pm 4.17$ ** | $9.03 \pm 4.37$ ** | $-0.41 \pm 2.41$ N.S | $7.14 \pm 3.65$ ** | $9.16 \pm 3.78$ ** |
| $Bladder\ V_{45Gy}$ (cc) | $52.83 \pm 36.40$ ** | $64.25 \pm 40.10$ ** | $4.90 \pm 51.97$ N.S | $48.39 \pm 41.08$ ** | $59.85 \pm 46.23$ ** |
| $Bowel\ V_{40Gy}$ (cc) | $47.96 \pm 44.03$ ** | $74.62 \pm 54.64$ ** | $-7.54 \pm 65.98$ N.S | $59.96 \pm 43.79$ ** | $85.95 \pm 46.67$ ** |
| $Bowel\ V_{45Gy}$ (cc) | $36.51 \pm 34.26$ ** | $56.62 \pm 41.31$ ** | $-2.42 \pm 47.82$ N.S | $41.45 \pm 33.79$ * | $61.89 \pm 34.84$ ** |
| $Rectum\ D_{mean}$ (Gy) | $9.68 \pm 2.75$ ** | $12.22 \pm 3.08$ ** | $-0.50 \pm 2.82$ N.S | $11.2 \pm .726$ ** | $13.17 \pm 2.91$ ** |
| $Rectum\ V_{35Gy}$ (cc) | $19.46 \pm 6.47$ ** | $22.22 \pm 7.79$ ** | $-10.87 \pm 16.44$ N.S | $26.9\ 8.15$ ** | $32.01 \pm 9.405$ ** |
| $Rectum\ V_{40Gy}$ (cc) | $21.28 \pm 7.02$ ** | $25.90 \pm 8.39$ ** | $-9.49 \pm 15.62$ N.S | $28.6 \pm 9.24$ ** | $33.59\ 10.59$ ** |

**Table 1** P-values of the Wilcox ranked-sign test performed between different Initial and Daily plans. The Bonferroni correction method was used to adjust these p values for type I errors. Significance is denoted:  *** p<0.001, ** p <0.01, * p ≤0.05, and N.S if p.>0.05.

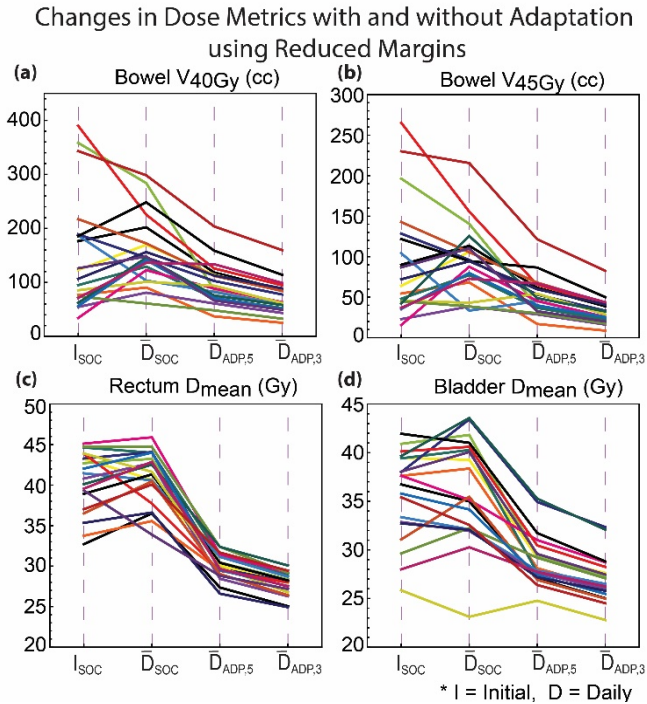

**Figure 2** Line plots of the (a) $Bowel\ V_{40Gy}$, (b) $Bowel\ V_{45Gy}$, (c) $Rectum\ D_{mean}$, and (d) $Bladder\ D_{mean}$ values from four of the plans: $Initial_{SOC}$, $\overline{Daily}_{SOC}$, $\overline{Daily}_{ADP,5mm}$, and $\overline{Daily}_{ADP,3mm}$. "I" and "D" denote Initial and Daily on the y-axis. Adapted plans on average significantly (p<0.05) decreased the dose metrics compared to the DailySOC plans.

In Figure 3, we have plotted the values for $Bowel\ V_{40Gy}$ from the $Daily_{SOC}$, $Daily_{ADP,5mm}$, $Daily_{ADP,3mm}$ plans from the 5 simulated treatments per patient as boxplots. We observed large intra- and inter-patient variability using the SOC non-adaptive plans. Adapting and reducing margins to either 5mm or 3mm always resulted in decreases to the mean $Bowel\ V_{40Gy}$ values, while the exact magnitude of the decrease was patient-specific.

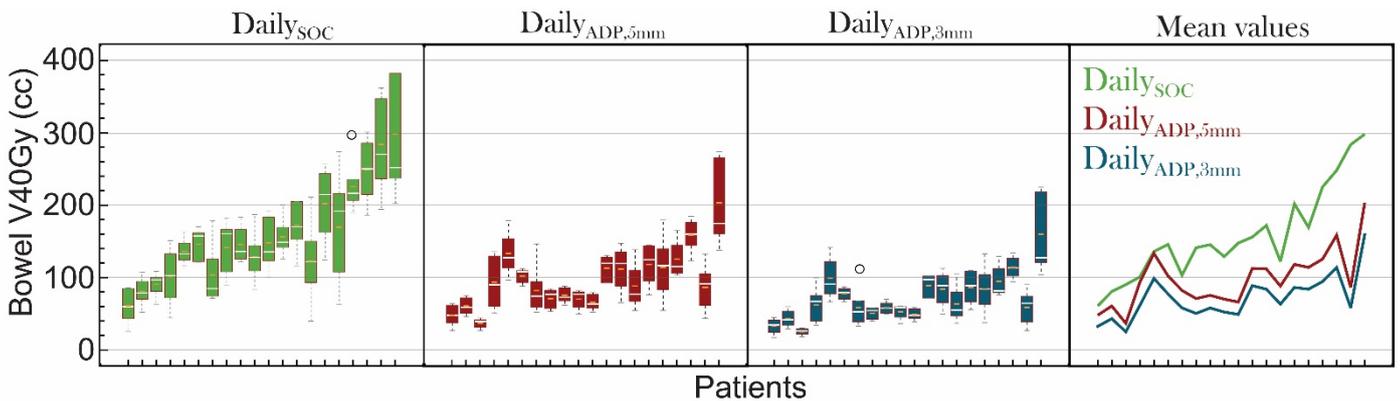

**Figure 3** Boxplot of the five fraction daily plan values for $Bowel\ V_{40Gy}$ (cc) for each patient for $Daily_{SOC}$, $Daily_{ADP,5mm}$, and $Daily_{ADP,3mm}$. Additionally, the mean values are plotted in the far right box for each patient from each type of daily plan. Reducing CTV-to-PTV margins to both 5mm and 3mm combined with daily adaptation reduced the average value for every patient although the magnitude was patient-specific.

### 3.2 Univariate Relationships

The R-squared between the difference in all the dose metrics of the initial plans ($Initial_{SOC-margin}$) and the difference in the four dose metric benefits on the daily plans ($\Delta_{SOC-ADP,margin}$) are plotted in **Figure 4** as a radar plot for (a) 3mm adaptive plans and (b) 5mm adaptive plans. The $Initial_{SOC-margin}$ metrics are on the axes and are the independent variable, while the relationship with the $\Delta_{SOC-ADP,margin}$ metrics (dependent variable) are represented by different colours. Relationships between the initial structure volumes and the overlap volumes with the PTVs versus the dose metrics of the daily plans were also studied for correlation with the benefits. The supplementary **S. Figure 2** shows that none of these volumes were strongly related, with most volume metrics having R-squared<0.3 for both PTV margins.

**Figure 4**b shows that the most predictive metric difference for most metrics was the difference in the same metric on the initial scan (e.g $\Delta_{SOC-ADP,margin}$ *Rectum $D_{mean}$ was only correlated to $Initial_{SOC-margin}$ Rectum $D_{mean}$*; none of the other Rectum metrics were correlated). An exception was $\Delta_{SOC-ADP,margin}$ Bowel $V_{40Gy}$ which had a strong relationship with $Initial_{SOC-margin}$ *Bowel $V_{40Gy}$*(cc) (R²=0.77 for 3mm, R²=0.80 for 5mm) and a moderate relationship with $Initial_{SOC-margin}$ *Bowel $V_{45Gy}$* (R²= 0.66 for 3mm and R²=0.56 for 5mm). Also $\Delta_{SOC-ADP,margin}$ *Bowel $V_{45Gy}$* had only moderate correlations overall and the highest of these was with the $Initial_{SOC-margin}$ *Bowel $V_{40Gy}$* (R²=0.42 for 3mm, R²=0.38 for 5mm).

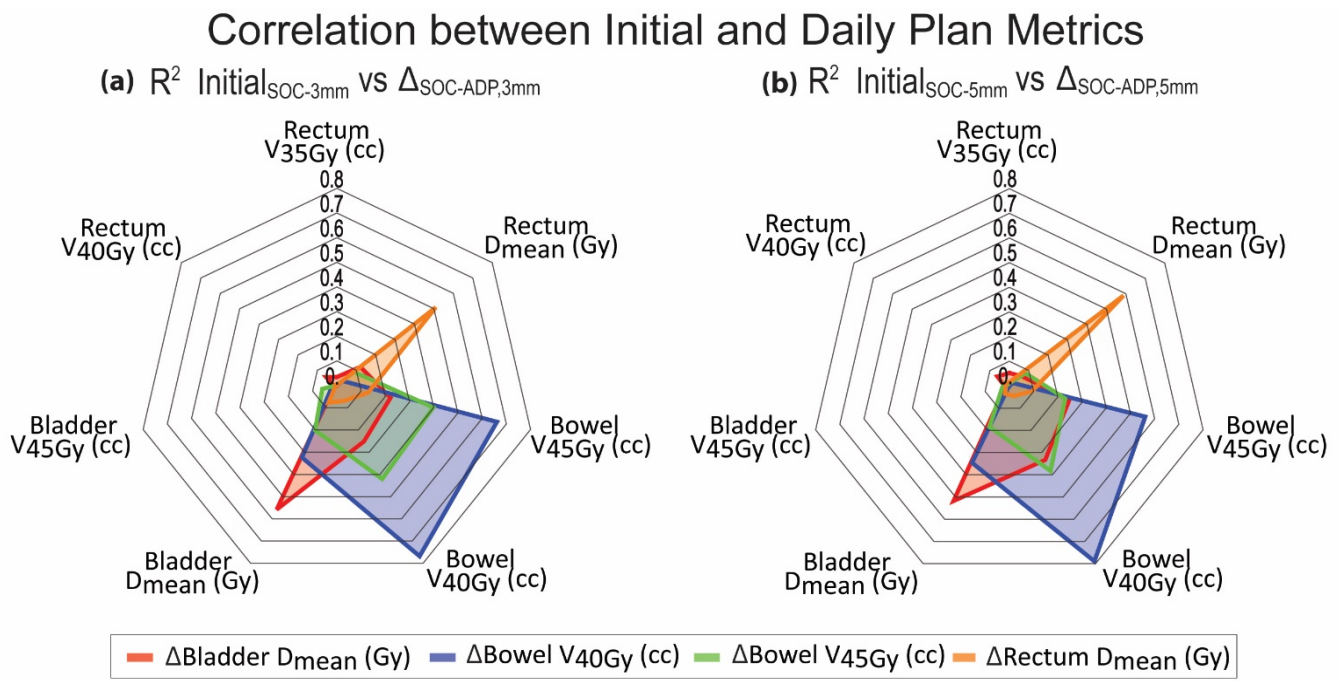

**Figure 4** Radar plots of the R² values for correlations between $Initial_{SOC-margin}$ on the axes and $\Delta_{SOC-ADP,margin}$ represented by different colours. The strongest relationships for the 3mm PTV were seen between **Bowel $V_{40Gy}$** (cc) $Initial_{SOC-3mm}$ and both $Bowel\ V_{40Gy}$(cc) $\Delta_{SOC-ADP,3mm}$ (blue) and $Bowel\ V_{45Gy}$(cc) (cc) $\Delta_{SOC-ADP,3mm}$ (green).

To clarify the association of the adapted dose benefits ($\Delta_{SOC-ADP,margin}$) with the initial plan metrics ($Initial_{SOC-margin}$), we plotted the $\Delta_{SOC-ADP,3mm}$ versus $Initial_{SOC-3mm}$ values with a linear best fit line in red (**Figure 5**) for the four predicted metric benefits against their same initial plan value differences. The dashed line is a $y = x$ line for reference. The highest R-squared value for the linear regression line was 0.77 for $Bowel\ V_{40Gy}$ (cc). **Figure 6** shows the same univariate relationships as **Figure 5**, but uses the data from the 5mm PTV Ethos plans. The overall relationships were very similar to the 3mm PTV adapted benefits and the highest R-squared of 0.8 was again for $Bowel\ V_{40Gy}$.

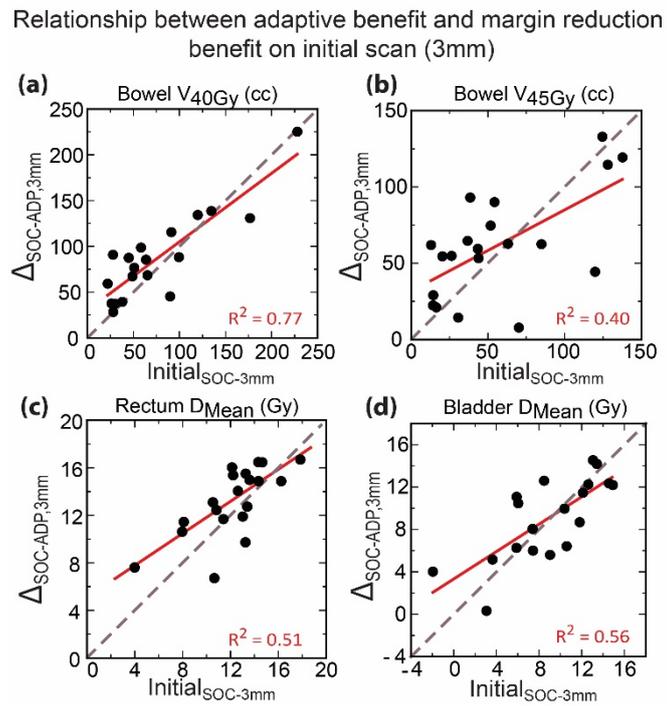

**Figure 5** Scatter plot of the adaptive benefit: $\Delta_{SOC-ADP,3mm}$ (along y-axis) versus margin benefit estimate from the initial plan: $Initial_{SOC-3mm}$ (along x-axis) for the PTV margin of 3mm. (a)-(d) represent the plots for the four different metrics: $Bowel\ V_{40Gy}$ (cc), $Bowel\ V_{45Gy}$ (cc), $Rectum\ D_{mean}$ (Gy) and $Bladder\ D_{mean}$ (Gy). The solid red lines represent the linear regression fit which yielded the $R^2$ given on the right bottom corner of the figures whereas, the brown dashed line is the $y = x$ line provided for reference. The strongest correlation was seen for predicting the adaptive benefit to the $Bowel\ V_{40Gy}$ (cc) using measurements from the initial plans.

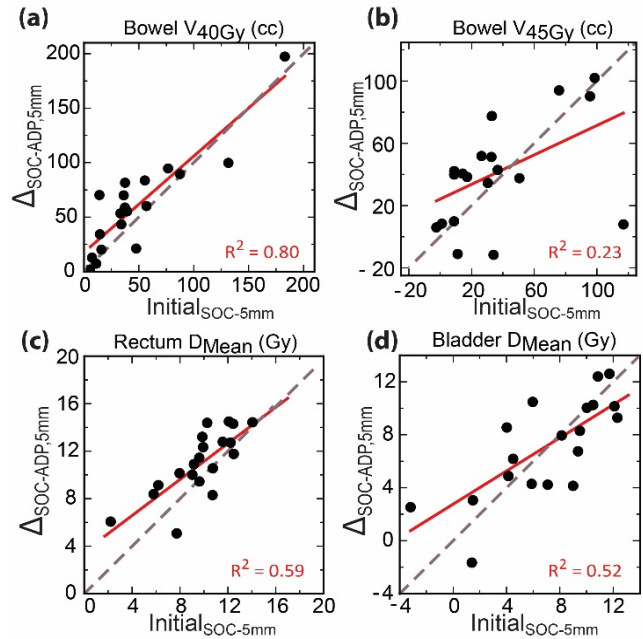

Relationship between adaptive benefit and margin reduction benefit on initial scan (5mm)

**Figure 6** Scatter plot of the adaptive benefit: $\Delta_{SOC-ADP,5mm}$ (along y-axis) versus margin benefit estimate from the initial plan: $Initial_{SOC-5mm}$ (along x-axis) for the PTV margin of 5mm. (a)-(d) represent the plots for the four different metrics: $Bowel\ V_{40Gy}$ (cc), $Bowel\ V_{45Gy}$ (cc), $Rectum\ D_{mean}$ (Gy) and $Bladder\ D_{mean}$ (Gy). The solid red lines represent the linear regression fit which yielded R² given on the right bottom corner of the figures whereas, the brown dashed line is the $y = x$ line provided for reference. A strong correlation was seen at this margin for predicting the adaptive benefit to the Bowel V40 from the measurements on the initial plans.

### 3.4 Multivariate Modelling

The univariate relationships indicated that more than one metric was associated with $Bowel\ V_{40Gy}$ (cc). To determine the most significant predictors we implemented a multivariate linear regression using the LASSO[35] technique. Potential covariates included any of the dose or volume metrics that had an R-squared> 0.2 in the univariate analysis. **Figure 7** shows the L1 regularization for the (a) 3mm and (b) 5mm margin plans respectively. For both margin schemas, we found that the most significant predictor metrics were the $Initial_{SOC-margin}\ Bowel\ V_{40Gy}$ (cc), $Initial_{SOC-margin}\ Bladder\ D_{mean}$ (Gy) and the Bladder Volume from the Initial CT scan.

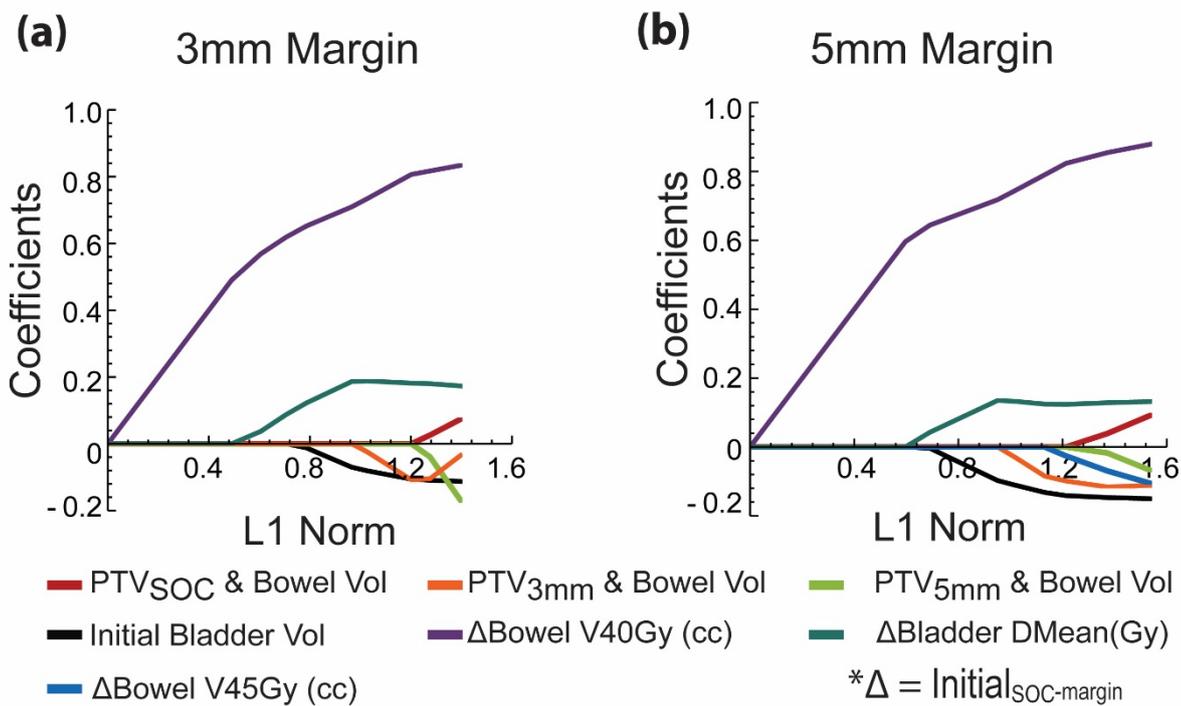

**Figure 7** L1 regularization performed for the 11 different predictors given in the legend for (a) 3mm PTV margin and (b) 5mm PTV margin. The three most predictive factors in both cases were $Initial_{SOC-margin}$ $Bowel\ V_{40Gy}$ (cc), $Initial_{SOC-margin}$ $Bladder\ D_{mean}$(Gy), and the bladder volume on the initial CT.

Using these covariates, we constructed a model to predict the $Bowel\ V_{40Gy}$ (cc) for each patient. We took a group of 19 of 20 patients to fit the coefficients for the multivariate model and tested it on the left-out patient to obtain a prediction for them where they did not bias the model fitting. This process was repeated leaving each patient out in turn and generating a prediction for them. Figure 8 shows the results of the model for the PTV margins of (a) 3mm and (b) 5mm. The true values are the observed values of each patient and predicted values are the values obtained from the model constructed over the other 19 patients. The resulting R-squared values were 0.656 and 0.663 for the 3mm and 5mm PTV models respectively. Visually the model predictions also appeared very similar for both PTV schemas.

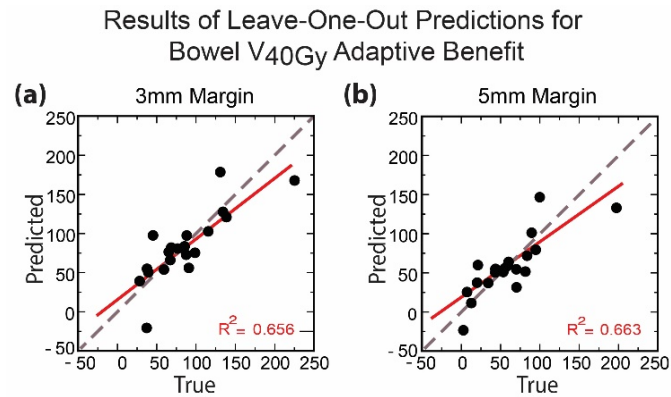

**Figure 8** Scatter plot of the predicted $\Delta_{SOC-ADP,margin}$ benefit to **Bowel $V_{40Gy}$** (along y-axis), from leave-one-out models using three significant metrics obtained by LASSO regularization as covariates, with respect to the true **Bowel $V_{40Gy}$** benefit: $\Delta_{SOC-ADP,margin}$ (along x-axis). The solid red line represents the linear regression fit line which yield $R^2$ given on the right bottom corner of the figures whereas, brown dashed line is the $y = x$ line provided for reference. For both the (a) 3mm and (b) 5mm margin PTV, the multivariate models had an intermediate $R^2$ >0.65.

### Discussion

This article aimed to predict the patient-specific dosimetric benefit of adaptive radiotherapy in a group of 20 locally advanced cervical cancer patients with intact uteri treated to 45Gy. Since bowel dose is related to many acute gastrointestinal complications including anorexia, nausea, diarrhoea etc., we focused our study on predicting the dosimetric benefits to *Bowel $V_{40Gy}$* (cc). We also included bladder and rectum metrics to evaluate their potential correlation.

The results showed that the patients with the highest benefit from ART (defined as relatively larger decreases in *Bowel $V_{40Gy}$* (cc)) were strongly correlated with those who had larger decreases in their bowel metrics from margin reduction on their initial plan. Additionally, the multivariate model was able to predict with reasonable accuracy the amount of reduction in *Bowel $V_{40Gy}$* (cc) a patient could expect with adaptation using either 3 or 5mm PTV margins. These successes are likely because the majority of the adapted benefit stems from the ability to use reduced target margins. Thus the impact of the reduced margins can be directly measured from the initial plan by reoptimizing on a smaller PTV. This result is beneficial for physicians and clinics who are looking for a quick way to evaluate if a patient would obtain substantial benefit from adaptation and thus direct their resources. The simplicity of the high-performing univariate model particularly lends itself to rapid clinical translation. A clinical team could estimate the benefit for their recent patients and set a threshold at which to recommend adapting based on the fraction of patients that they have capacity to adapt. Interestingly, we did not find any evidence that this benefit could be predicted by looking at even simpler geometric measures such as the volumes of the structures or the volume overlaps with the different PTV margins.

In the univariate analysis, we also observed that benefits in *Rectum $D_{mean}$* (Gy) and *Bladder $D_{mean}$*(Gy), had R-squared values above but close to 0.5 meaning only slightly more than half the variance in this variable is explained by the impact of margin reduction. This means additional information would be required to predict the average benefit from adaptation. The remaining variance could be due to variations in organ filling and daily positioning which cannot be measured from the initial scan. However, it is possible the magnitudes of these variations could be measured early in treatment such as from the first week's imaging and used to inform a model to predict benefit from adapting the remaining 4-5 weeks of treatment. This data could also improve our existing positive results for *Bowel $V_{40Gy}$*.

In this analysis we examined two different reduced PTV margins for daily adaptation: 3 and 5mm. So far, a recent paper evaluated intrafraction motion during the online CBCT adaptive process for cervical cancer patients and found that 5mm margins led to 98.39 +/- 3% of the CTV receiving 100% of the dose[36]. Unsurprisingly, the greatest factor impacting CTV coverage was the time required to perform CTV edits, with longer times leading to greater changes in bladder volume and thus worse daily CTV coverage. Thus the 5mm data in this study represents reasonable margins for a new adaptive program using this same technology, while 3mm represents a possible further target reduction for patients that demonstrate no intrafraction motion or for adaptive procedures that are faster due to improvements in technology (e.g. improved auto-segmentation, reduced optimization time). For both margin schemas, our results for the univariate and multivariate models were very similar. As a result, clinics could prioritize the same patients as highest benefit regardless of which amount of margin reduction they felt was the best match for their workflow.

This is the first paper to attempt to idenfity patients with proportionally higher benefit from online ART [37]. A similar approach was evaluated for offline adaptation in head and neck cancer patients and found that the benefit to the parotids of a single offline adaptive plan mid-way through treatment could be predicted from the initial plan Parotid $D_{mean}$ (Gy)[38]. Additionally, groups have investigated methods to predict when daily replanning is needed on a per-fraction basis in order to reduce the total number of fractions being adapted per patient[37]. Accurate prediction of benefit from high-resource activities are critical for both clinics and patients. The extra time spent planning each adaptive fraction can double or triple the total time patients spend in clinic and thus is not negligible. Additionally, this procedure generally requires extra oversight from physicians and physicists, and thus reduces the time they have available for other existing clinical duties.

While we focused on predicting $Bowel\ V_{40Gy}$ (cc) for cervical cancer patients, the same approach could be used in other disease sites and our multivariate modelling approach could also be used with a binary patient-classification using multiple metrics. For example, a high-benefit patient could be identified as one with an anticipated 30% decrease in any OAR metric with adaptation. Furthermore, while our analysis used CBCT-guided daily ART, the results are agnostic to daily imaging type and would apply to clinics using MRI guided ART.

Overall, this analysis demonstrates an approach of building a model to predict patient-specific benefit from ART which could eventually be used to triage patients for this highly resource-intensive technology. One of the weaknesses of the study is that the data was averaged over the five fractions to compare plans and, thus the true accumulated dose for the entire course was not considered. This was due to the challenges in accurate dose accumulation in radiotherapy particularly for structures without distinctive landmarks such as the bowel[39]. That being said, using the mean dose value over multiple adaptive treatments spread throughout a course of radiotherapy presents a reasonably conservative estimate of the dosimetric benefit to the patients.

This analysis was also limited by the small study population of 20 patients and the homogenous nature of the cohort. This approach allowed us to identify patients within this cohort that were most likely to benefit, but it is possible that a different cervical cohort (e.g. node positive) could have a higher overall benefit than our node-negative cohort. In future studies, we will incorporate more patients including those who are post-hysterectomy, node-positive, or are receiving different prescription regimens. Each of these characteristics can then also be set as a potential covariate in the model-building process and used to predict the highest benefit cervical patients.

In summary, our data suggest that there is a noticeable correlation between the initial plan values and the final dose benefits from adaptive treatment for the $Bowel\ V_{40Gy}$. As this metric is correlated to acute GI toxicity, early prediction of the benefit will help clinicians prioritize patients for resource-intensive daily ART. We expect these results to further improve as we incorporate greater variation in patient types. Successful modelling of adaptive benefits for cohorts of interest will help us implement this new technology more effectively in clinics particularly in clinics who are resource constrained.

**Supplementary Data**

| Metrics | $Initial_{SOC}$ | | $\overline{Daily}_{SOC}$ | | $Initial_{5mm}$ | | $Initial_{3mm}$ | | $\overline{Daily}_{ADP,5mm}$ | | $\overline{Daily}_{ADP,3mm}$ | |
|---|---|---|---|---|---|---|---|---|---|---|---|---|
| | μ | σ | μ | σ | μ | σ | μ | σ | μ | σ | μ | σ |
| $Bladder\ D_{mean}$(Gy) | 35.7 | 4.5 | 36.1 | 5.2 | 28.8 | 2.2 | 26.7 | 2.2 | 29.0 | 2.7 | 27.0 | 2.3 |
| $Bladder\ V_{45Gy}$ (cc) | 79.0 | 48.1 | 74.1 | 55.8 | 26.2 | 25.0 | 14.7 | 19.2 | 25.7 | 18.0 | 14.2 | 11.3 |
| $Bowel\ V_{40Gy}$ (cc) | 150.5 | 106.2 | 158.1 | 64.8 | 102.6 | 70.8 | 75.9 | 57.6 | 98.1 | 39.1 | 72.1 | 30.9 |
| $Bowel\ V_{45Gy}$(cc) | 92.4 | 70.0 | 94.8 | 43.9 | 55.9 | 45.8 | 35.8 | 34.8 | 53.4 | 23.3 | 32.9 | 16.1 |
| $Rectum\ D_{mean}$ (Gy) | 40.5 | 3.8 | 41.0 | 3.3 | 30.8 | 2.3 | 28.2 | 2.3 | 30.0 | 1.5 | 27.8 | 1.4 |
| $Rectum\ V_{35Gy}$ (cc) | 38.4 | 13.0 | 49.3 | 16.0 | 18.9 | 7.7 | 14.2 | 6.3 | 22.4 | 9.0 | 17.2 | 7.6 |
| $Rectum\ V_{40Gy}$ (cc) | 34.9 | 11.7 | 44.4 | 14.9 | 13.6 | 6.0 | 9.0 | 4.5 | 15.8 | 6.5 | 10.8 | 5.0 |

**S. Table 1:** Mean and standard deviation from the 20 patients for each dose metric from each of the plans in this analysis.

| Metrics | p-values | | | | |
| --- | --- | --- | --- | --- | --- |
| | $Initial_{SOC}$ vs $Initial_{5mm}$ | $Initial_{SOC}$ vs $Initial_{3mm}$ | $\dfrac{Initial_{SOC}}{\text{vs } Daily_{SOC}}$ | $\dfrac{\overline{Daily_{SOC}}}{\text{vs } \overline{Daily_{ADP,5mm}}}$ | $\dfrac{\overline{Daily_{SOC}}}{\text{vs } Daily_{ADP,3mm}}$ |
| $Bladder\ D_{mean}$ (Gy) | 0.0053 | 0.0039 | 1.0000 | 0.0039 | 0.0033 |
| $Bladder\ V_{45Gy}$ (cc) | 0.0033 | 0.0033 | 1.0000 | 0.0033 | 0.0033 |
| $Bowel\ V_{40Gy}$ (cc) | 0.0033 | 0.0033 | 1.0000 | 0.0033 | 0.0033 |
| $Bowel\ V_{45Gy}$ (cc) | 0.0045 | 0.0033 | 1.0000 | 0.017 | 0.0033 |
| $Rectum\ D_{mean}$ (Gy) | 0.0033 | 0.0033 | 1.0000 | 0.0033 | 0.0033 |
| $Rectum\ V_{35Gy}$ (cc) | 0.0033 | 0.0033 | 1.0000 | 0.0033 | 0.0033 |
| $Rectum\ V_{40Gy}$ (cc) | 0.0033 | 0.0033 | 1.0000 | 0.0033 | 0.0033 |

**S. Table 2:** P-values of the Wilcox ranked-sign test performed between different Initial and Daily plans. The Bonferroni correction method was used to adjust these p values for type I errors.

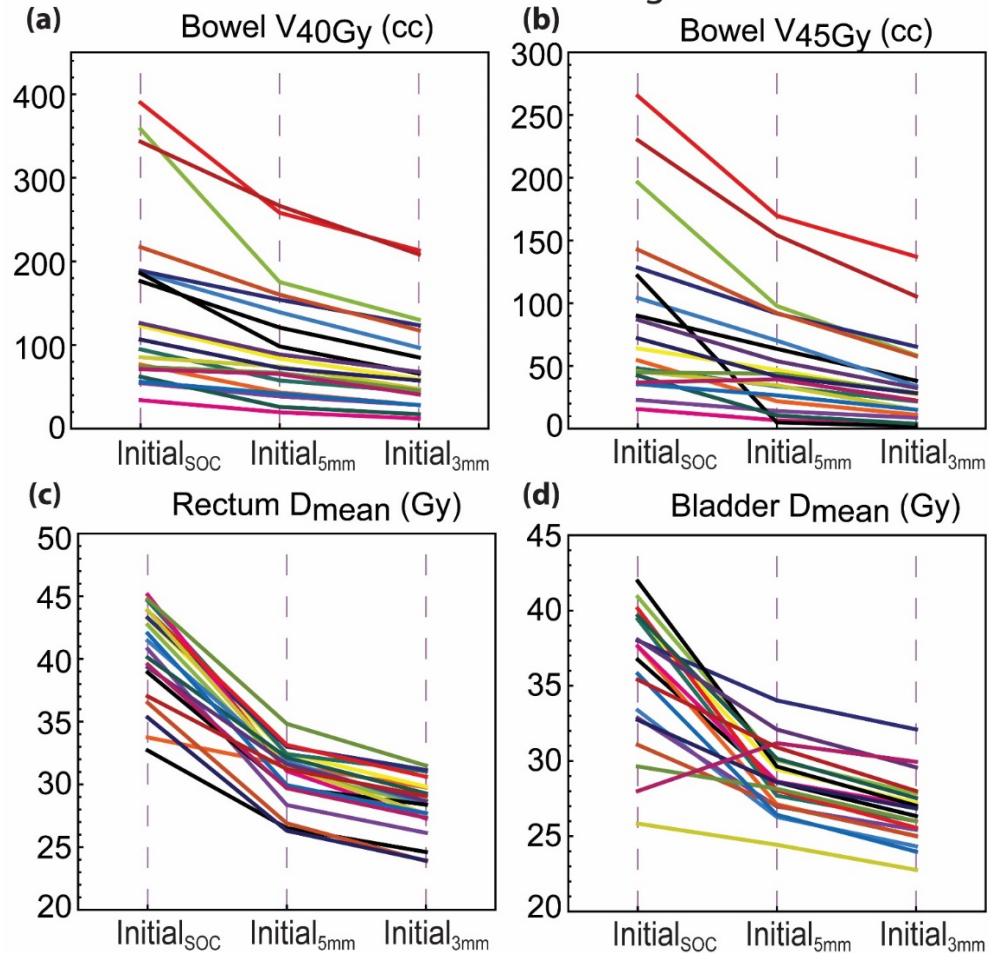

**S. Figure 1:** Line plots of the (a) $Bowel\ V_{40Gy}$, (b) $Bowel\ V_{45Gy}$, (c) $Rectum\ D_{mean}$, and (d) $Bladder\ D_{mean}$ values from the three initial plans: $Initial_{SOC}$, $Initial_{5mm}$, $Initial_{3mm}$. Reducing the margins on average significantly (p<0.05) decreased the dose metrics compared to the $Initial_{SOC}$ plans.

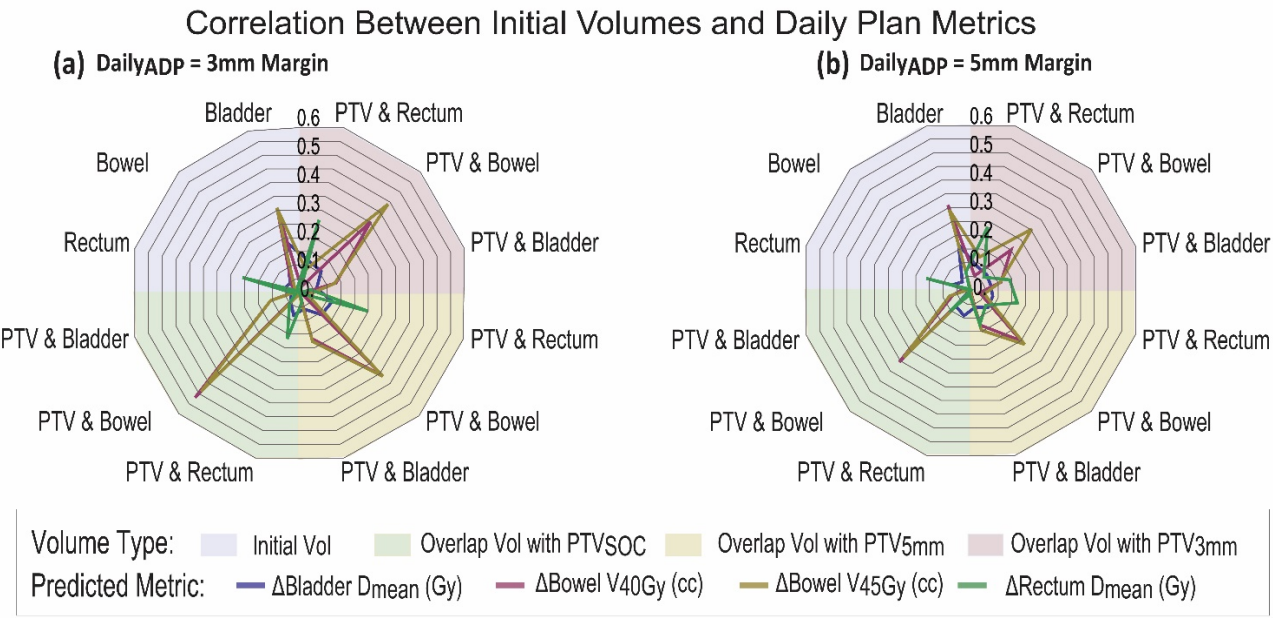

**S. Figure 2** Radar plots of the R$^2$ values for correlations between the volume and volume overlap metrics on the axes and $\Delta_{SOC-ADP,margin}$ adaptive dose benefits represented by different colours. For both margins, no strong relationships (R-squared>0.7) were observed. However, the largest values were seen for the overlaps between the different PTVs and the Bowel volume for both the 3 and 5mm PTVs and the adaptive benefit to the bowel: $\boldsymbol{Bowel\ V_{40Gy}}$(cc) $\boldsymbol{\Delta_{SOC-ADP,margin}}$ (pink) and $\boldsymbol{Bowel\ V_{45Gy}}$(cc) $\boldsymbol{\Delta_{SOC-ADP,margin}}$ (yellow).